\documentclass[lettersize,journal]{IEEEtran}
\usepackage{cite}
\usepackage{amsmath,amssymb,amsfonts}
\usepackage{algorithmic}
\usepackage{graphicx}
\usepackage{textcomp}
\usepackage{xcolor}
\usepackage{fancyhdr}
\usepackage{hyperref}

\usepackage[ruled,vlined]{algorithm2e}
\usepackage{epsfig}
\usepackage{bm}
\usepackage{multirow}
\usepackage{mathtools}

\usepackage{array}
\usepackage[caption=false,font=normalsize,labelfont=sf,textfont=sf]{subfig}
\usepackage{stfloats}
\usepackage{url}
\usepackage{verbatim}

\newcommand{\squeezeup}{\vspace{0mm}}
\newcommand{\squeezeupless}{\vspace{0mm}}

\newcommand{\Design}{FHEmem\xspace}

\begin{document}

\title{\Design: A Processing In-Memory Accelerator for Fully Homomorphic Encryption}

\author{Minxuan Zhou$\dagger$, Yujin Nam$\dagger$, Pranav Gangwar$\dagger$, Weihong Xu$\dagger$, Arpan Dutta$\dagger$, 
Kartikeyan Subramanyam$\dagger$\\ Chris Wilkerson$\ddagger$, Rosario Cammarota$\ddagger$, Saransh Gupta$\diamond$, and Tajana Rosing$\dagger$\\
University of California San Diego$\dagger$, Intel Labs$\ddagger$, IBM Research$\diamond$\\
\{miz087,yujinnam,pgangwar,wexu,adutta,kasubram,tajana\}@ucsd.edu$\dagger$,\\ \{chris.wilkerson,rosario.cammarota\}@intel.com$\ddagger$, saransh@ibm.com$\diamond$}

\maketitle

\begin{abstract}
Fully Homomorphic Encryption (FHE) is a technique that allows arbitrary computations to be performed on encrypted data without the need for decryption, making it ideal for securing many emerging applications. However, FHE computation is significantly slower than computation on plain data due to the increase in data size after encryption. Processing In-Memory (PIM) is a promising technology that can accelerate data-intensive workloads with extensive parallelism. However, FHE is challenging for PIM acceleration due to the long-bitwidth multiplications and complex data movements involved. 
We propose a PIM-based FHE accelerator, \Design, which exploits a novel processing in-memory architecture to achieve high-throughput and efficient acceleration for FHE. We propose an optimized end-to-end processing flow, from low-level hardware processing to high-level application mapping, that fully exploits the high throughput of \Design hardware.
Our evaluation shows \Design achieves significant speedup and efficiency improvement over state-of-the-art FHE accelerators.
\end{abstract}


\section{Introduction}
The data explosion leads to an increasing trend of cloud-based outsourcing. Extensive outsourcing significantly increases the risk of sensitive data leaking, necessitating data encryption for protection.
Fully homomorphic encryption (FHE) is an emerging technology that enables computations on encrypted data without user interference~\cite{tfhe1,ckks,gentry09,bgv}.
FHE provides end-to-end data security during the outsourcing, including data transfer and computation, without any requirements for the underlying system and hardware.
However, FHE is several orders of magnitude slower than plain data while requiring a large memory footprint~\cite{bts, craterlake,ark,sharp,memfhe}.
The inefficiency of FHE results from the data and computation explosion after encryption. Even though FHE can encrypt a vector into a single ciphertext~\cite{ckks,bgv}, the ciphertext size is still large and includes two or more high-degree polynomials (e.g., $2^{17}$) with long-bit coefficients (e.g., $>$1000 bits). 

\begin{figure}[t!]
    \centering
    \epsfig{file=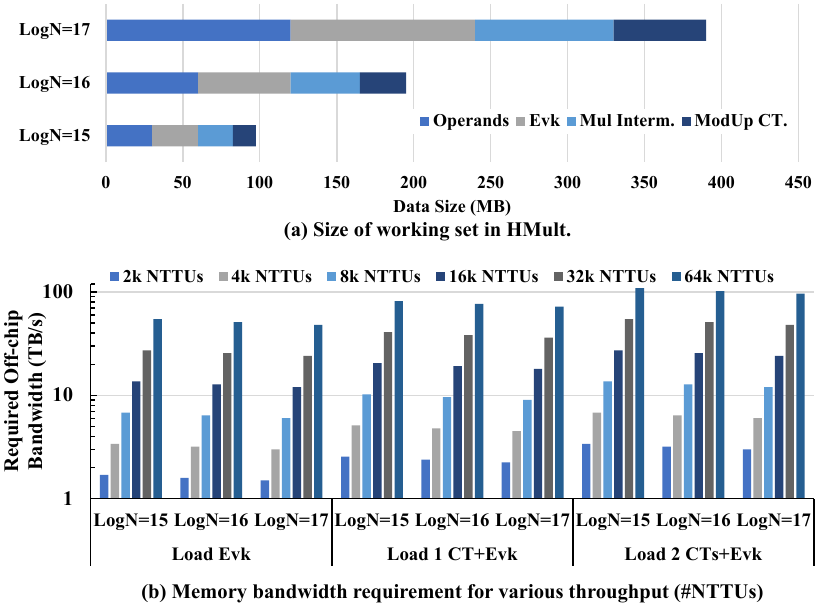, width=\columnwidth}
    \squeezeup
    \squeezeup
    \caption{The memory bandwidth requirements when varying the on-chip throughput (\#NTTUs). We assume L=30, LogQ=1920.}
    \squeezeup
    \squeezeup
    \label{fig:mem_bound_motiv}
\end{figure}

Such issues motivate researchers to develop customized accelerators that provide 4 orders of magnitude speedup over conventional systems~\cite{bts, craterlake,ark,sharp,memfhe}. However, existing accelerators are still significantly bounded by the data movement even with large and costly on-chip scratchpads~\cite{bts,sharp,craterlake}. 
As shown in Figure~\ref{fig:mem_bound_motiv}(a), each homomorphic multiplication ($HMul$) requires $98MB$ to $390MB$ working set for $LogN=15$ to $LogN=17$. In Figure~\ref{fig:mem_bound_motiv}(b), we follow the method in previous work~\cite{bts} to analyze the memory bandwidth required by different numbers of number theory transform units (NTTUs) under 3 data loading scenarios during a holomorphic operation with key-switching operation (KSO) which is the most expensive FHE primitive. 
Our investigation shows simply doubling the throughput of existing accelerators (1K to 2K) may require over 3TB/s of off-chip bandwidth. Previous accelerators adopt large on-chip storage, up to 512MB, to hold the large working set of FHE computation. However, large on-chip storage can still suffer from frequent off-chip data transfers due to cache misses on extensive FHE data.
Therefore, it is challenging to achieve both high compute throughput and high memory bandwidth on conventional architectures for FHE applications.

In this work, we exploit the processing in memory (PIM) acceleration for FHE, which is promising to support extensive parallelism with high internal memory bandwidth~\cite{drisa,floatpim, neuralcache,computedram,simdram,transpim,memfhe}. 
There are several types of PIM architectures that support PIM in different levels in the memory architecture, including subarray-level~\cite{simdram,drisa,computedram}, bank-level~\cite{fimdram_stack,sk_aim}, and channel-level~\cite{tesseract}. These PIM technologies adopt high-throughput processing elements that fully exploit the internal memory links that provide higher bandwidth than the off-chip data path.
Even though the high parallelism and bandwidth of PIM potentially fit the data-intensive parallel computations of FHE, there exist several challenges that existing PIM-based architectures cannot easily solve.
First, FHE works on high-degree polynomials with long-bit coefficients and is multiplication-intensive. Such long-bit multiplication is challenging to all existing PIM technologies. For near-bank PIM, the throughput is limited by bank IO width. Furthermore, even though the bank-level PIM can adopt highly efficient multipliers, reading the data from the memory cells is still energy-consuming. The bit-serial subarray-level PIM can exploit a significantly large number of internal links (e.g., bitlines). However, the number of operations required by subarray-level PIM may increase quadratically with the operand bit-length. These energy-consuming operations run in a lock-step manner, significantly increasing delay and energy.
Our evaluation shows such PIM technologies fail to provide comparable throughput and energy efficiency to the state-of-the-art FHE accelerators (Section~\ref{sec:pim_motiv}). Integrating more functionality in the sense amplifiers can solve the problem, but introducing significant modifications in conventional DRAM structures~\cite{drisa}.
The second challenge of PIM acceleration for FHE is the complex data movement patterns, including the base conversion and number theoretic transform (NTT), that the existing memory architecture cannot efficiently support.

To tackle these challenges, we propose \Design, an accelerator based on a novel high-bandwidth memory (HBM) architecture optimized for the efficiency of processing FHE operations. 
\Design introduces a novel near-mat processing architecture, which integrates compute logic near each mat without changing the area-optimized mat architecture. \Design exploits the existing intra-memory data links, with careful extensions based on the practicability, to enable efficient in-memory processing of various challenging FHE operations. Furthermore, we propose a software-level framework to map FHE programs onto \Design hardware. We propose a load-save pipeline that can fully utilize the memory for FHE programs to support high-throughput computing with minimum data loading overhead.

We summarize the contributions of this work as follows:
\begin{itemize}
    \item We propose an FHE accelerator with a novel near-mat processing that supports high-throughput and energy-efficient in-memory operations. Our design efficiently exploits the existing data paths with practical modifications in DRAM that introduce relatively lightweight overhead compared to prior high-throughput PIM solutions. 
    \item We propose an FHE mapping framework that generates a load-save pipeline and data layout that maximizes the utilization of \Design for general FHE programs.
    \item We rigorously explore and evaluate different design dimensions of \Design to balance the performance, energy efficiency, and chip area. As compared to state-of-the-art FHE ASICs~\cite{sharp,craterlake}, \Design achieves 4.0× speedup and 6.9× efficiency improvement.
\end{itemize}
\section{Background and Motivation} \label{sec:motivation}

\subsection{Fully Homomorphic Encryption}\label{sec:fhe}
We focus on CKKS scheme~\cite{ckks} which is widely used in many application domains because it supports real numbers and  SIMD packing~\cite{han2019logistic,resnet,lola}. 

\textbf{Basics of CKKS:}
We define the polynomial ring $R=\mathbb{Z}[X]/(X^N+1)$, where $N$ is power of 2. We denote $R_q=R/qR$ for residue ring of $R$ modulo an integer $q$. The ecurity parameter $\lambda$ sets the ring size $N$ and a ciphertext modulus $Q$. For each plaintext message, $m(X)$, encryption \texttt{Encrypt}($m(X),s(X)$) generates a ciphertext $c = (b(X), a(X))$, where $b(X) = a(X) \cdot s(X) + m(X) + e(X)$, $a(X)$ is uniformly sampled from $R_{Q}$, and $s(X)$ and $e(X)$ are sampled from a key/error distribution respectively. Each ciphertext can pack up to $N/2$ real numbers to support SIMD processing on all packed numbers~\cite{bgv, ckks}. The original modulus $Q=\sum_{l=1}^{L}{q_l}$ of a ciphertext decreases with homomorphic multiplications by a rescaling process that reduces the modulus by a $q_l$ each time.
Therefore, CKKS is a leveled homomorphic scheme that only supports $L$ levels of multiplications for each ciphertext. 
The technique to recover the ciphertext level is bootstrapping~\cite{ckks_boot}.

\textbf{Arithmetic Operation:}
Given two ciphertexts $c0,c1 \in R^{2}_{q_l}$, where $q_l$ is the modulus at level $l$, the polynomial operation can homomorphically evaluate the arithmetic for plaintexts. The homomorphic multiplication (\texttt{HMul}) between two ciphertexts is complex: $c0*c1=(I_0, I_1, I_2)=(c0_{a}c1_{a}, c0_{a}c1_{b}+c1_{a}c0_{b}, c0_{b}c1_{b}) \in R^{3}_{ql}$, where $I_2$ is encrypted under the secret key $s^2$. This requires a re-linearization operation with an expensive key-switching process on $I_2$ (Section~\ref{sec:key_switch}). 
The rescaling is applied during the relinearlization, using the divide and round operation: \texttt{ReScale}$(C)=\lfloor{\frac{q_{l-1}}{q_{l}}C}\rceil(mod\ q_{l-1})$ to rescale the ciphertext as well as the modulus.

\textbf{Rotation:}
CKKS supports homomorphic rotation which rotates the plaintext vector by an arbitrary step. The rotation is implemented by Galois group automorphism~\cite{automorphism} which consists of mapping on each coefficient $a_i$: 
$\sigma_{k}(a_i) \xrightarrow[]{} (-1)^sa_{ik\ mod\ N}$
,where $k$ is an odd integer satisfying $|k|<N$ and $s=0$ if $ik\ mod\ 2N<N$ ($s=1$ otherwise). Each automorphism $\sigma_k$ implements a \texttt{Rotate}$(\delta)$ which rotates the plaintext by $\delta$. Each automorphism also requires a \texttt{KeySwitch} after each \texttt{Rotate}.

\textbf{NTT and Residue Number System:}
Number Theoretic Transform (\texttt{NTT}) is a widely used technique to optimize polynomial multiplications.
\texttt{NTT} transforms two input polynomials of a multiplication, $a$ and $b$, to \texttt{NTT(a)} and \texttt{NTT(b)}. We can calculate $NTT(a*b) = NTT(a) \odot NTT(b)$, where $\odot$ denotes element-wise multiplication with $O(N)$ complexity, where $N$ is the polynomial degree. An inverse NTT (\texttt{iNTT}) can transform $NTT(a*b)$. The complexity of \texttt{NTT} and \texttt{iNTT} is $O(NlogN)$, faster than the original polynomial multiplication with $O(N^2)$ complexity.
Residue number system (RNS) is a technique that avoids computation on large values. We adopt the full-RNS version of CKKS~\cite{full_rns}. 
For polynomials in $R_Q$, the scheme chooses a set of pair-wise coprime integers $q_i$ where $i \in [0,L)$ and $q_0q_1...q_L=Q$. Each polynomial $a$ is represented by $L$ polynomials $a[0...L]$, where $a[i] \in R_{q_i}$. We can evaluate \texttt{FUNC}$(a,b)$ by independently calculating \texttt{FUNC}$(a[i],b[i])$ based on Chinese Remainder Theorem. These RNS moduli are also used as the leveled modulus, so each multiplication removes one RNS polynomial in the ciphertext.

\textbf{Key Switching:}\label{sec:key_switch}
Key switching is the most expensive high-level operation in CKKS, where we use the state-of-the-art generalized key switching algorithm~\cite{ckks_boot}. 
The key of key switching is the multiplication between the input ciphertext $c$ and the evaluation key $evk$.
However, the naive multiplication will cause the overflow on modulus $Q=q_0q_1...qL$. 
To avoid overflow on modulus $Q=q_0q_1...qL$, $evk$ has a larger modulus $PQ$ with the special modulus $P=p_0p_1...p_k$. Thus the first step is to convert $c$ with modulus $Q$ into a ciphertext with $PQ$ by a base conversion (\texttt{BConv}):

\squeezeup
\squeezeup
\begin{equation} \label{eq:bconv}
\squeezeupless
BConv_{Q,P}(a_Q)=([\sum_{j=0}^{L}{[a[j]*\hat{q_j}^{-1}]_{q_j}*\hat{q_j}}]_{p_i})_{0\leq i<k}
\end{equation}

\texttt{BConv} requires the data in the original coefficient domain. We need to apply an \texttt{iNTT} on the data before \texttt{BConv}. \texttt{BConv} features an all-to-all reduction between different $q_j$ and $p_i$ residual polynomials. 
We convert the \texttt{BConv} result back to the \texttt{NTT} domain to efficiently process the multiplication with $evk$. The algorithm converts the result with modulus $PQ$ back to modulus $Q$ using \texttt{BConv}. Recent advanced CKKS schemes exploit a configurable $dnum$ value to factorize the modulus $Q$ into $dnum$ moduli to increase higher multiplication level~\cite{ckks_boot}. 

\subsection{Memory Issues of FHE Accelerators}
FHE features large polynomial operations and complex data dependency caused by \texttt{(i)NTT} and \texttt{BConv}.
Recent works~\cite{f1,bts, craterlake,ark,sharp,memfhe} have proposed customized accelerators for FHE. Even though these accelerators achieve up to 4 orders of magnitude speedup over CPUs, they suffer from limited memory bandwidth. For example, previous work~\cite{bts} observed that the excessive usage of high-throughput function units might be a waste - it would be cost-efficient to determine the throughput of on-chip processing elements based on the available memory bandwidth.
Such memory issues result from the large data size required for each FHE operation. Existing FHE accelerators adopt large on-chip storage (180MB for SHARP~\cite{sharp}, 256MB for CraterLake~\cite{craterlake}, and 512MB for BTS~\cite{bts}/ARK~\cite{ark}) to reduce the frequent off-chip data loading. However, such large on-chip storage may still be insufficient for FHE, as shown in Figure~\ref{fig:mem_bound_motiv}. For large FHE parameter settings, on-chip storage may only store working sets of one or two $HMul$, leading to frequent off-chip data loading when locality is low. The analysis in Figure~\ref{fig:mem_bound_motiv}(b) shows 2k NTTUs require at least 1.5TB/s when only loading $evk$, and the bandwidth requirement goes up to 3TB/s when the accelerator needs to load both $evk$ and two operands. 3TB/s is expected to require 3 HBM3 stacks~\cite{hbm3}. If we increase the throughput to 64k NTTUs, which can fully parallelize operations for $LogN=17$, the bandwidth requirement can be as high as $100TB/s$. Considering it is challenging to significantly increase either the memory bandwidth or the on-chip storage, processing in memory can be a promising alternative.

\subsection{In-DRAM PIM Technologies}~\label{sec:dram}
\begin{figure}[t!]
    \centering
    \epsfig{file=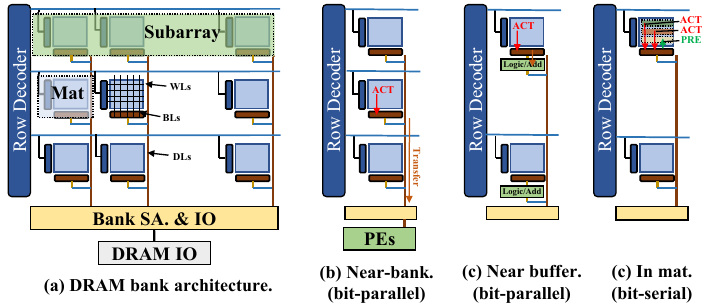, width=\columnwidth}
    \squeezeup
    \squeezeup
    \caption{Different in-DRAM PIM technologies.}
    \squeezeup
    \squeezeup
    \label{fig:dram_pim}
\end{figure}
This work focuses on DRAM-based PIM technologies which support larger capacity than SRAM~\cite{neuralcache} and lower latency than non-volatile memories~\cite{floatpim}.
Figure~\ref{fig:dram_pim}(a) shows a DRAM bank, which is the basic hardware component in DRAM. A bank consists of 2D cell arrays and peripherals to transfer data between DRAM cells and IOs. The memory cells are grouped into subarrays, each consisting of a row of mats. Each mat has local sense amplifiers (row buffers) sensing a horizontal wordline (WL) through a set of vertical bitlines (BLs). Sense amplifiers in mats of a subarray form the subarray row buffer. 
Upon receiving access, the bank activates corresponding WL in subarray row buffer and transfers the whole WL to bank-level sense amplifiers via data lines (DLs).

Several DRAM-based PIM technologies support operations in different levels of DRAM architecture. The first technology is near-bank processing, which integrates processing elements (e.g., vector ALUs, RISC-V processor, etc.) near the bank SA and IO~\cite{fimdram_stack}. Each bank-level PE is customized to fully utilize the data link bandwidth for processing. PEs in different banks run in parallel to fully utilize the internal bandwidth in DRAM. The second type of PIM augments the subarray sense amplifiers (row buffers) with compute logic~\cite{drisa}. Due to the constrained chip area, the near-buffer PIM only adopts logic gates, full adders, and shift circuits for multi-bit operations. Compared to near-bank processing, near-buffer PIM supports wider input (e.g., 8192b vs. 256b) and can exploit the subarray-level parallelism~\cite{salp}. These two types of PIM work on the data with a horizontal layout where each data is stored across multiple BLs in a WL. The third type of PIM uses a vertical bit-serial scheme that lays out each data in different WLs of a BL~\cite{simdram}. The bit-serial PIM directly generates the result of computation between different WLs by exploiting the charge-sharing effect of the DRAM mechanism. Such in-mat bit-serial processing does not introduce significant modifications in DRAM. However, the bit-serial computation is slow and power-consuming where an n-bit multiplication using the bit-serial PIM requires around $7n^2$ DRAM activations for 8k values.

\subsection{Challenges of FHE acceleration using PIM}~\label{sec:pim_motiv}
Even though existing PIM solutions can exploit the large internal DRAM bandwidth for high-throughput computation, FHE is still extremely challenging for PIM acceleration.

\begin{figure}[t!]
    \centering
    \epsfig{file=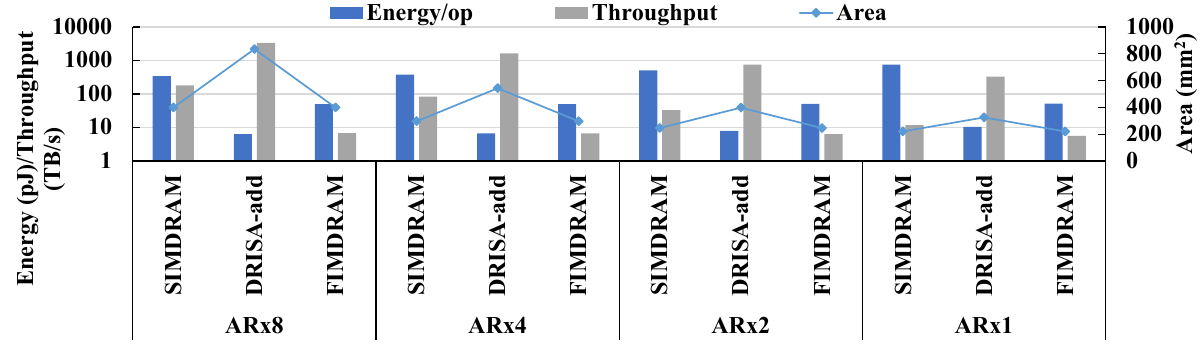, width=\columnwidth}
    \squeezeup
    \squeezeup
    \caption{Throughput and energy efficiency of 32-bit multiplication using different PIM technologies (32GB).}
    \squeezeup
    \squeezeup
    \label{fig:pim_motiv}
\end{figure}
\subsubsection{Long-bit multiplication}
As introduced in Section~\ref{sec:fhe}, the basic data structure of FHE is high-degree polynomial, whose coefficient can exceed $2^{1000}$. The RNS-decomposed polynomials still have at least 28-bit coefficients due to the limitation of modulus selection~\cite{craterlake,ckks}. 
As the complexity of PIM multiplication significantly increases as the bit-length increases, PIM (especially bit-serial in-mat) may suffer from long latency and high energy efficiency due to costly row activations and precharges. One way to improve the performance and energy efficiency of PIM is by increasing the aspect-ratio (AR) of DRAM mat~\cite{reduce}. A high AR mat has fewer WLs (rows) and shorter BLs than a low AR mat. Shorter BLs can significantly reduce the latency and energy of activation and precharge. For instance, ARx4 mat (128 rows) has half the cycle and consumes 33\% less energy than ARx1 mat (512 rows)~\cite{reduce,drisa}. Furthermore, increasing the AR also increases the number of subarrays in a bank, leading to a higher degree of parallelism. The downside of high AR is the large area overhead caused by more sense amplifiers and peripherals.

Figure~\ref{fig:pim_motiv} shows the throughput and energy efficiency of different PIM technologies for 32-bit multiplications on a 32GB HBM2E-based architecture (Section~\ref{sec:exp}). We evaluate three existing PIM architectures,  FIMDRAM~\cite{fimdram_stack}, DRISA~\cite{drisa}, and SIMDRAM~\cite{simdram}, that represent near-bank, near-buffer, and in-mat bit-serial PIM respectively.
The result shows FIMDRAM and SIMDRAM provide 6.8TB/sand 180.6TB/s throughput while consuming 49.8pJ and 342.9pJ energy for each operation using ARx8 memory.  
As a reference, the recent FHE accelerator~\cite{craterlake}, which adopts 150k 28b multipliers, can provide 1PB/s of peak throughput while consuming only 4.1pJ for each multiplication, indicating both FIMDRAM and SIMDRAM are not promising for FHE. DRISA~\cite{drisa} provides over 3PB/s throughput and consumes 6.32pJ for each operation in ARx8 architecture. However, DRISA~\cite{drisa} requires a significant change in the DRAM architecture, incurring around 100\% area overhead in high-AR architectures. Furthermore, manufacturing DRISA has significant challenges as the modified sense amplifiers cannot easily be aligned with area-optimized bitlines. To tackle these challenges, \Design adopts a novel near-mat processing that integrates compute logic near mat while keeping the mat structure intact, incurring less area overhead than DRISA. Even though the theoretical throughput and energy efficiency of \Design are lower than DRISA, our experiments show that higher throughput may not effectively improve the performance due to the data movement (Section~\ref{sec:exp}). Overall, \Design is a more efficient processing paradigm than prior PIM solutions.

\subsubsection{Data Transfer Patterns}
Another critical challenge of FHE for PIM is the variety of data transfer patterns in different FHE operations. Specifically, the \texttt{BConv} requires the data movement between different RNS polynomials, followed by coefficient-wise operations; \texttt{(i)NTT} and automorphism require data permutation across coefficients of each RNS polynomial. 
Unfortunately, the conventional memory IO cannot efficiently handle such complex data movement patterns. 
For \texttt{BConv}, each output RNS polynomial has dependencies with all input RNS polynomials. As each RNS polynomial is large (e.g., 512KB for LogN=64 with 64-bit coefficients), we must distribute RNS polynomials over different memory banks. In conventional memory, such inter-bank data movements take up the shared bus of each channel, leading to significant data movement overhead. For \texttt{(i)NTT} and automorphism, the data movement exhibits a fine-grained pattern within a polynomial. For PIM processing, the coefficient-wise permutation requires permutations between BLs which is not supported in the current memory architecture.
\Design supports these FHE-specific data transfers efficiently at a relatively low cost by exploiting existing intra-memory data links and adding additional links to the less-dense metal layer in DRAM, without introducing complex permutation networks.
\section{\Design Hardware Architecture}\label{sec:hardware}

\begin{figure}[t!]
    \centering
    \epsfig{file=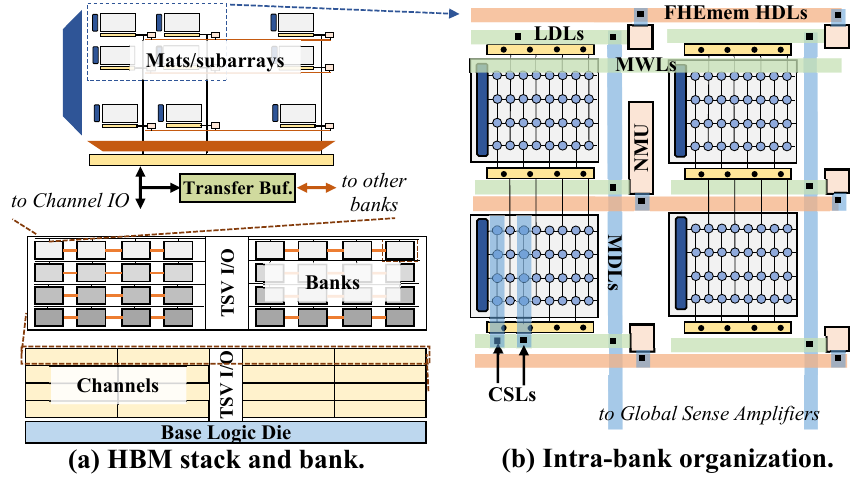, width=\columnwidth}
    \caption{The hardware architecture of \Design.}
    \squeezeup
    \label{fig:pim_arch}
\end{figure}

The high-level architecture of \Design is based on high-bandwidth memory (HBM), as shown in Figure~\ref{fig:pim_arch}. Specifically, each HBM stack consists of one base die and multiple DRAM dies in a 3D structure. All DRAM dies are divided into multiple channels, each connecting to DRAM part through an independent set of through silicon vias (TSVs). Each channel consists of several banks, and the detailed internal structure of the bank is introduced in Section~\ref{sec:dram}.
\Design adopts a new near-mat PIM architecture that modifies the DRAM bank architecture to support high-throughput computations while utilizing available DRAM internal links for various FHE operations. The key customized components of \Design include near-mat units (NMUs), horizontal data links, and inter-bank connection.

\subsection{Near-mat unit}
\begin{figure}[t!]
    \centering
    \epsfig{file=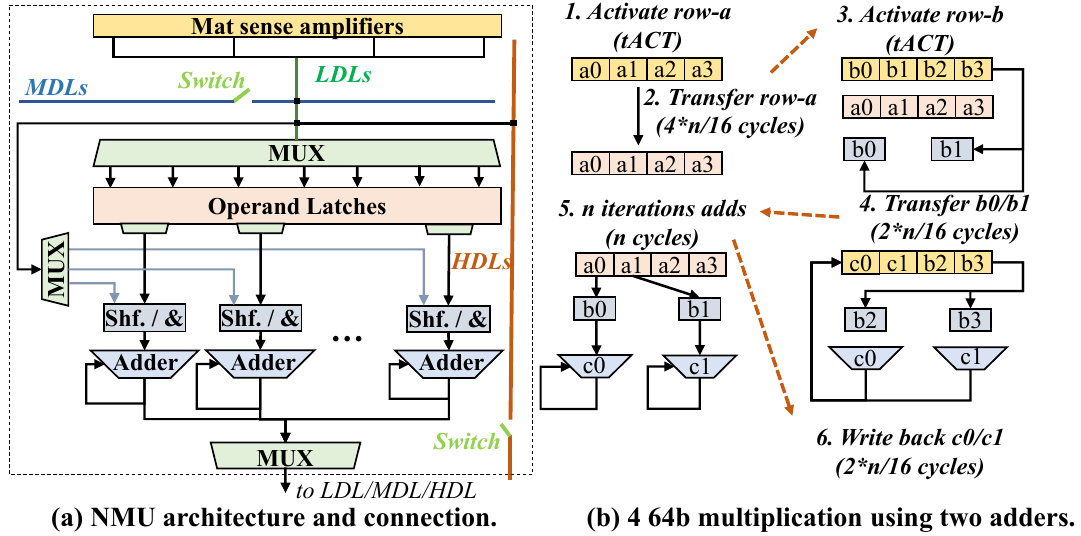, width=\columnwidth}
    \caption{NMU architecture and NMU-based PIM.}
    \squeezeup
    \label{fig:adder}
\end{figure}
\Design supports in-memory computations by connecting each mat to a near-mat unit (NMU) via the local data lines (LDLs). Each NMU consists of full adders, shifters, AND logic, and latches to compute FHE operations, shown in Figure~\ref{fig:adder}(a). In addition to the mat sense amplifiers, NMU can also receive data from other NMUs via inter-NMU connection. The design of NMU differs from previous PIM-logic integration, DRISA~\cite{drisa}, in three ways. First, NMU places all customized logic outside the mat, which is optimized for area efficiency. On the contrary, DRISA~\cite{drisa} integrates logic and latches with the mat sense amplifiers and bitlines, which may incur significant changes to the area-optimized mat. Second, DRISA~\cite{drisa} integrates logic to all bitlines in a mat, causing a large area overhead while only gaining moderate performance benefits due to the unbalanced compute and data movement~\cite{drisa}. In \Design, we explore the processing throughput of NMU under different architecture configurations and observe the most efficient design does not adopt the maximum throughput (Section~\ref{sec:exp}). Third, NMU in \Design can support permutations required by FHE using multiplexers in the data path.

To compute FHE arithmetic (i.e., modular arithmetic), NMU requires several steps in mat, data links, and NMU. Figure~\ref{fig:adder}(b) shows an example of processing 4 64b multiplications in an NMU with 2 64b adders. For generality, we denote each mat row can store N n-bit values (N=4 and n=64 in this example), and NMU has M n-bit adders. First, the mat must activate an operand row (1) and transfer the row to the row-size operand latches (2). Next, the mat activates the second operand row (3) and transfers M-value blocks to the shifter and AND logic (4). When both operands are ready, the shifter and AND logic will generate a partial product using the second operand (b0 and b1) and bit masks of a specific bit of the first operand (a0 and a1 in the latches). NMU takes n cycles to compute an n-bit multiplication (5). After processing an M-value block, NMU writes the result back to the mat (6) and loads the next M-value block for processing (7). Like DRISA~\cite{drisa}, NMU only needs two row activations for each vector processing but requires serial data transfers via LDLs. NMU can serially write back values in a different order to support permutation.

\subsection{Inter-NMU Connection}
\begin{figure}[t!]
    \centering
    \epsfig{file=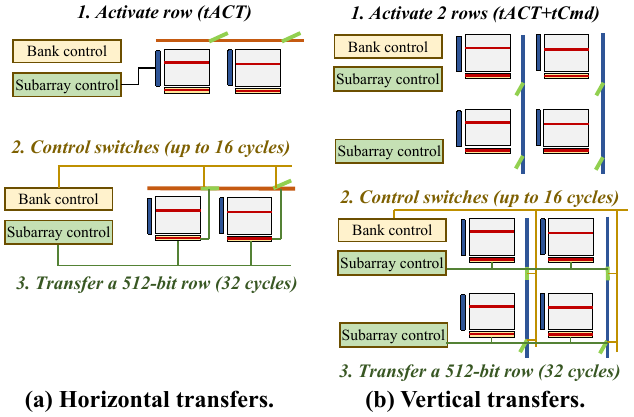, width=0.8\columnwidth}
    \caption{Steps of data transfer in \Design.}
    \squeezeup
    \label{fig:transfer_rebuttal}
\end{figure}
To efficiently process various FHE data transfer patterns, \Design enables data transfer between NMUs in horizontal and vertical directions. In the vertical direction, \Design utilizes the master data lines (MDLs) which connect all NMUs in the same mat column. In the horizontal direction, \Design adds extra data links, horizontal data links (HDLs), to each subarray (a row of mat). The HDLs in each subarray support the same bitwidth as the MDLs in each mat column (i.e., 16-bit). For both directional links, we add small isolation transistors (switch in Figure~\ref{fig:adder})~\cite{lisa} to serve as switches that can disconnect each link at a certain point. NMUs separated by the off switches can transfer data independently, significantly improving the bandwidth for intra-bank and intra-subarray data movements. 

In \Design, each horizontal and vertical inter-NMU connection transfers a 512-bit mat row to another mat, where Figure~\ref{fig:transfer_rebuttal} shows the detailed steps. Before transferring, \Design activates the participating row(s) in subarray(s). The horizontal (vertical) transfer requires the activation in 1 (2) subarray(s). \Design then turns on or off the switches based on the transfer pattern (Section~\ref{sec:control}). To reduce the area and energy cost, \Design uses a single control signal for all switches in a row for vertical links (MDLs), similar to LISA~\cite{lisa}; for horizontal links (HDLs), \Design adopts a single control signal for all switches in a column. The bank-level logic has the bookkeeping logic for switches' states. Setting up switches requires up to 16 cycles because we map each polynomial to a $16\times 16$ mat array (Section~\ref{sec:bank_layout}). Last, the subarray controller controls the near-mat peripheral to send/receive data based on the transfer pattern. Transferring 512-bit data via 16-bit data links requires 32 cycles.

\subsection{Inter-bank Connection}~\label{sec:conn}
As introduced in Section~\ref{sec:fhe}, FHE has dependencies between RNS polynomials of a ciphertext during \texttt{BConv}. To store a ciphertext, which can consist of tens of RNS polynomials, we need to distribute these RNS polynomials over multiple banks, and each BConv requires a large amount of inter-bank data transfers because of an all-to-all dependency.
For such inter-bank data movements, a naive way is to use the conventional channel-level data bus. 
Unfortunately, such centralized data movements fail to match the throughput of in-memory computations, significantly slowing down the acceleration. 
Furthermore, the cost of fully-connected interconnect can be prohibitively high. To support the inter-bank communication with satisfactory performance, while introducing reasonable overhead in HBM, we propose a partial chain interconnect network between banks inside a channel. 
The partial chain network connects neighboring banks in each bank group, using 256b-wide transfer links and per-bank transfer buffers. Specifically, the transfer buffer in each bank can communicate with the local master data lines (MDLs) and the transfer links to the neighboring banks. We add two 256b transfer buffers in each bank to support seamless transfers between banks. The customized links and buffers enable parallel inter-bank data transfers across different banks in a channel, avoiding sequential transfers via the original channel IO. When transferring a whole row between two neighboring banks, the source bank drives 256b data blocks from the selected subarray SA via MDLs to the transfer buffer, which sends data blocks to the transfer buffer of the destination bank via either the customized links (neighboring banks) or the original channel IO (non-neighboring banks). The destination bank writes data blocks to the selected subarray SA via MDLs. 

\subsection{\Design Controllers}~\label{sec:control}

\begin{table}[t!]
  \centering
  \caption{\Design NMU commands}

  \label{tab:inst}
\resizebox{\columnwidth}{!}{
\begin{tabular}{l|l|l}
\hline
\textbf{Command}             & \textbf{Description}                               & \textbf{Cycles}                             \\ \hline
nmu\_ld            & load data from SA column address to NMU latches           &     $size / 16$    \\
nmu\_st             & store from NMU latch to SA column address          &  $size / 16$  \\
nmu\_hmov           & horizontal data movement between NMUs in a subarray      &   $size / 16$  \\
nmu\_vmov           & vertical data movement between subarrays  & $size / 16$   \\
nmu\_add & Start addition using selected latches with or without shift \&AND & $\#shifts$ \\
nmu\_pst        & store different NMU latches (64-bit) to SA column address & 4 \\ \hline
\end{tabular}
}
\end{table}

\begin{figure}[t!]
    \centering
    \epsfig{file=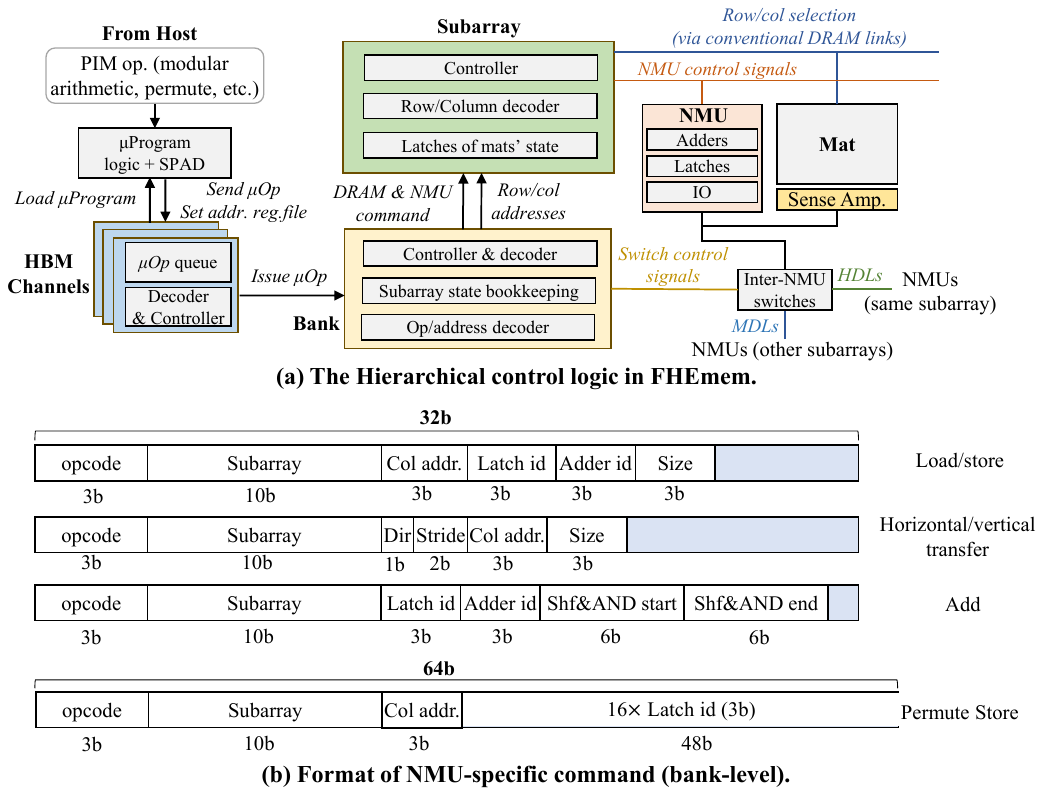, width=\columnwidth}
    \caption{The control logic of \Design.}
    \squeezeup
    \label{fig:control_rebuttal}
\end{figure}

\Design needs modifications in bank/subarray controllers to support its various functions.
 To support the in-memory computation, \Design exploits the existing control logic of SIMDRAM~\cite{simdram}. Figure~\ref{fig:control_rebuttal}(a) shows the processing flow of hierarchical control with the annotation of \Design extension. The host CPU sends \textit{bbop} instructions to each memory controller (i.e., channel-level controller in HBM). 
\Design requires an extension of \textit{bbop} instructions for modular arithmetic and permutation for FHE. Each memory controller has a micro-program control logic that translates each \textit{bbop} to a micro-program which is a sequence of DRAM subarray-level commands. 
Previous works~\cite{simdram} has shown the micro-program control logic only incurs negligible area overhead (less than $0.1mm^2$).

\Design extends the bank-level control to decode NMU commands and dispatches signals to subarray-level logic and the isolation transistors (switches). \Design requires the extra logic for subarray-level parallelism~\cite{salp}, including the bookkeeping logic in the bank controller for the status of all subarrays and the extra address latches in each subarray row decoder. The subarray-level control sends signals to all NMUs in the subarray based on the command.
Furthermore, \Design adds more latches in the subarray column decoder to support the permutation operations. The extra logic and control signals in \Design also introduce insignificant overhead (Table~\ref{tab:area_power}) because they are placed outside the mat.

\Design adds several new subarray-level commands for NMU processing, as shown in Table~\ref{tab:inst} and Figure~\ref{fig:control_rebuttal}(b). The subarray-level commands control the same behaviors of all mats/NMUs in a subarray, except $nmu\_pst$ which stores different latches in different mats back to SA (used for automorphism). 
For NMU loading and storing, \Design supports the flexibility of selecting columns in NMU latches, adder latches, and sense amplifiers, enabling permutation. The horizontal data movement has predefined patterns, defined by $direction$ and $stride$, to support NTT data movements. The vertical data movement has more flexibility to transfer data between two subarrays. The $add$ command in \Design indicates the latch-adder pair for computation and whether to use shift/AND and the bit position of shift/AND. To exploit subarray-level parallelism and minimize the command patching latency (i.e., minimize the number of commands), most \Design commands, except the permute store ($nmu\_pst$), can be configured for processing large data size.

Figure~\ref{fig:control_rebuttal}(b) shows the format of the 7 \Design commands (3-bit opcode). 
To support the precision requirement of FHE, each \Design command processes 64-bit data. Therefore, the column/latch address and the size of data of near-mat processing can be presented by 3 bits (i.e., 8 possible 64-bit data in 512-bit row in a mat). In our architectural exploration (Section~\ref{sec:exp}), each bank has 128 (ARx1) to 1024 (ARx8) subarrays, requiring 10-bit to denote the subarray. Each subarray contains 16 mats so we can use 3-bit mat id, 1-bit direction and  2-bit stride, to represent all horizontal movements in a subarray. The addition command includes a range of $shift\&AND$ for multiplication - both the start and end shift steps can be represented by 6-bit (i.e., up to 64 bits).
For the permute store command, a 48-bit field is reserved for the vector of latch addresses in 16 NMUs of a subarray. The issue time of 32-bit (64-bit) command to each bank via 16-bit command/address bus is 2 (4) cycles, respectively.

\subsection{Practicality of \Design}
Commodity DRAMs are optimized for cost so that DRAM process only adopts 3 metal layers, with 1 layer (M1) for bitlines (vertical), 1 layer (M2) for LDLs and master word lines (MWLs) (horizontal), and 1 layer (M3) for column select lines (CSLs) and MDLs~\cite{mem_scaling}. M1 layer is the only fine-pitch (low energy efficiency) layer, optimized for the area of bitlines in a mat. M2 and M3 support more energy-efficient wires with larger area overhead (i.e., 4x pitch of M1). 
The design of \Design considers the cost-efficiency of modifying commodity DRAM, avoiding the high-cost changes in cost-sensitive components.
In \Design, the additional horizontal data lines are placed in M2 which incurs less pressure in the routing. 
M3 (vertical) is more dense than M2 (horizontal) because both CSLs and MDLs on M3 extend across multiple mats while M2 only has MWLs shared by multiple mats.
Furthermore, HDLs connect NMUs, which are placed outside of dense mats (DRAM cell arrays). 
As illustrated in previous work~\cite{mem_scaling}, changes near the mat are the most costly, including bitline sense amplifiers, local wordline drivers, row logic, and column logic. Changes outside the subarray logic are relatively inexpensive because of the low-density logic blocks.
These characteristics make HDLs and NMUs more practical than previous PIM solution that integrates complex logic in bitline sense amplifiers (e.g., DRISA~\cite{drisa}) in conventional DRAM technology.

\section{Optimized Processing Flow of FHE in \Design}\label{sec:flow}
The proposed \Design hardware supports high-throughput computations and data movements. The next challenge is to efficiently utilize \Design for FHE applications. This section introduces an optimized end-to-end processing flow that determines the data layout and processing flow of \Design for FHE applications by adopting several algorithm and compiler-level optimizations.

\subsection{\Design Data Layout}~\label{sec:bank_layout}
\begin{figure}[t!]
    \centering
    \epsfig{file=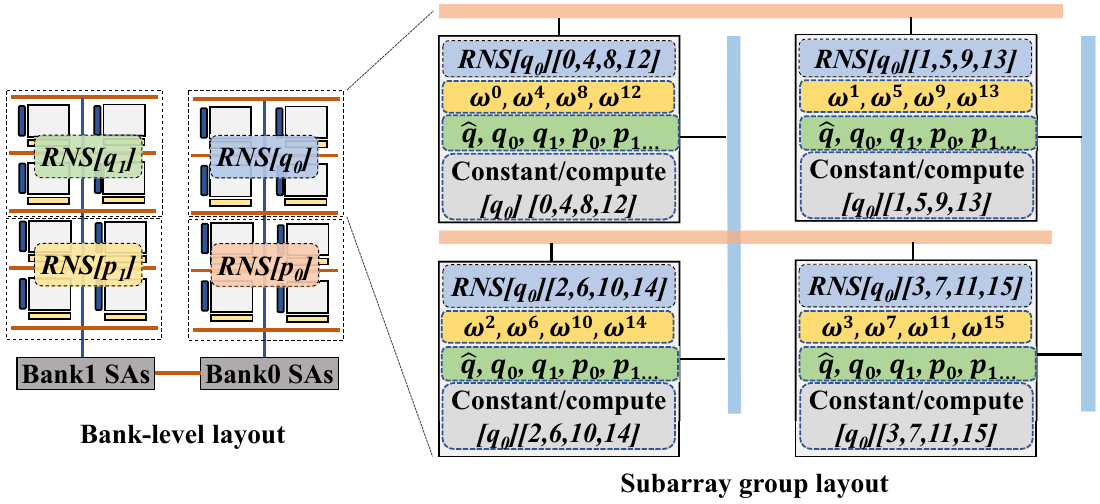, width=\columnwidth}
    \squeezeup
    \squeezeup
    \caption{The data layout in \Design.}
    \squeezeup
    \squeezeup
    \label{fig:bank_layout}
\end{figure}
In PIM acceleration, data layout is critical to determine the detailed computation and data movement. 
Figure~\ref{fig:bank_layout} shows the optimized data layout in \Design. Each ciphertext contains a group of RNS polynomials, each of which is a vector of $N$ $b$-bit integers. To exploit the high throughput of PIM, we distribute RNS polynomials of a ciphertext, including the original RNS terms ($RNS_{ql}$) and the special terms ($RNS_{pk}$) used for key switching, across multiple banks using a round-robin method. The figure shows an example of allocating two original RNS terms and two special RNS terms in a bank. 

\subsubsection{Layout in subarray groups}
We divide subarrays into subarray groups, basic memory partitions for polynomials. Specifically, each subarray group contains a continuous set of subarrays (e.g., 2 subarrays) which is a 2D array of mats (e.g., 2$\times$2). The 2D distribution allows \Design to balance the inter-mat data movements during various FHE operations (esp. NTT).
\Design distributes coefficients of a polynomial across the mat array in an interleaved way, similar to previous work~\cite{bts}, to efficiently support automorphism (Section~\ref{sec:automorphism}). In our setting, each subarray group contains 16 subarrays (16 $\times$ 16 mats), requiring each mat to use 32 rows to store 256 64-bit coefficients.

\subsubsection{Layout for computation}
For a computation using two RNS polynomials, we align both polynomials in the same column in a subarray. 
Each subarray group reserves rows for operand polynomials used for computation, including key-switching keys, constant polynomials, and other ciphertexts. If a subarray group computes with two polynomials in different subarray groups, the memory issues data movements that may happen inside a bank or across different banks. 

\subsubsection{Layout for constants}
In addition to RNS polynomial, we need to allocate rows for several constants for FHE operations, including \texttt{(i)NTT} twiddle factors, moduli, scaled inverse moduli, etc. To avoid duplicating twiddle factors across NTT steps that require large memory, we store the vector of twiddle factors which contains $N$-th roots of unity, in the same order as the polynomial coefficients. Before each (i)NTT stage $k$, \Design first dynamically computes the twiddle factor $\omega^{ik}$ for coefficient $i$ by multiplying the twiddle factor in the previous stage with $\omega^{i}$. For moduli, we keep one copy in each mat in the subarray group, so that each NMU can load the corresponding value independently during the computation.

\subsection{Algorithm-optimized modular reduction}~\label{sec:bit_opt}
A modular reduction follows each FHE arithmetic. We exploit the Montgomery algorithm~\cite{montgomery} requiring two multiplications, one addition, and one subtraction. NMUs in \Design processes n-bit multiplication using n serial additions. Therefore, we exploit algorithm optimization to significantly reduce the latency and energy for modular arithmetic.
Specifically, we select moduli that are friendly to serial computations while satisfying security requirements and (i)NTT. We exploit the moduli selection technique proposed in previous works~\cite{optimized_modulus_1} that select moduli has the form of $2^{b}\pm 2^{sh_1}\pm 2^{sh_2} \pm .. \pm 2^{sh_{h-2}} \pm 1$, where $h$ is called hamming weight. Using a modulus with a hamming weight of $h$, we only need to issue $h$ additions during the multiplication, hence reducing the addition steps from $n$ to $h$. 
The hamming weight optimization only applies to computations with constant, including the multiplication with modulus and the multiplication with reduction factor in Montgomery reduction. The advantage of Montgomery reduction over Barrett reduction~\cite{barrett} is that both reduction factor and modulus in Montgomery reduction can have a low hamming weight, and it only requires single bit-length computation. We note that prior FHE accelerators~\cite{craterlake} also adopted a similar optimization that customized the modular multiplier for Montgomery-friendly moduli. However, their modular multipliers cannot process computations using moduli with different characteristics (e.g., required by other applications). \Design provides more flexibility by using addition as the basic computing step.

\begin{figure}[t!]
    \centering
    \epsfig{file=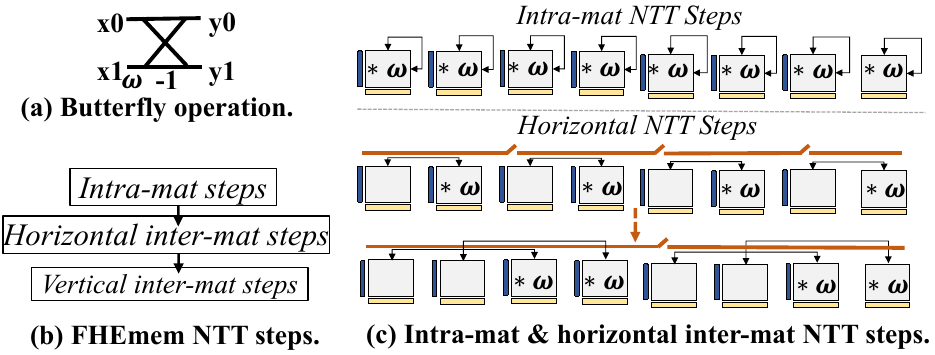, width=\columnwidth}
    \squeezeup
    \squeezeup
    \caption{NTT support in \Design.}
    \squeezeup
    \squeezeup
    \label{fig:ntt}
\end{figure}

\subsection{\Design NTT}~\label{sec:ntt}
\texttt{NTT} executes several steps of permutation and computations. Each \texttt{NTT} step requires several butterfly operations (Figure~\ref{fig:ntt}(a)) on pairs of coefficients, that multiply the coefficients with the twiddle factors, permute the coefficients, and then update the coefficients. 
\Design processes each (i)NTT operation in three stages, intra-mat, horizontal inter-mat, and vertical inter-mat, depending on the butterfly stride of each (i)NTT step (Figure~\ref{fig:ntt}(b)). 
For intra-mat steps, where coefficients of each butterfly operation are in the same mat, NMUs in a subarray group independently process computation and permutation. The horizontal inter-mat steps exchange the coefficients between mats in the same row, for which \Design uses HDLs for efficient data transfers, as shown in Figure~\ref{fig:ntt}(c). Specifically, \Design turns on/off the switches of NMUs on HDLs, where the connected segments can independently transfer data. 
Data transfers using the same connected segment are scheduled sequentially. Therefore, as the number of connected segments changes over (i)NTT stages, the transfer latency of different (i)NTT stages varies. The vertical inter-mat NTT steps are processed similarly to the horizontal steps but using MDLs. The key novelty of \Design on (i)NTT operations is that \Design does not introduce complex butterfly networks in the memory. Instead, \Design exploits the existing DRAM internal links (i.e., MDLs and LDLs) with efficient customizations (i.e., NMU, switches, HDLs).

\subsection{Base Conversion}~\label{sec:bconv_layout}
\textit{BConv} is a costly but frequent operation in FHE. 
As introduced in Section~\ref{sec:motivation}, to generate each special RNS polynomial of \texttt{BConv} (with modulus $pk$), each input RNS polynomial with modulus $qi$ first multiplies $[\hat{qi}^{-1}]_{qi}$ and $[\hat{qi}]_{pk}$. Such partial products are reduced to each special RNS polynomial $pk$. \Design parallelizes multiplications in different subarray groups with different input polynomials. To reduce the partial products, \Design first accumulates partial products in the same bank using NMUs and MDLs because partial products of different polynomials in a bank are aligned either in the same subarray group or in different subarray groups in the vertical direction. Therefore, the intra-bank accumulation can be processed in an adder-tree manner by exploiting switches in MDLs. \Design processes the final reduction in the bank of the output polynomial, requiring data transfers of partial products from all other banks. \Design handles such inter-bank data movements using the customized interconnection (Section~\ref{sec:conn}). To parallelize the computation, each bank processes different output polynomials simultaneously. \Design determines the optimized schedule based on the number of banks used for the ciphertext, the number of input/output RNS polynomials, and the underlying interconnect structure.

\subsection{Automorphism}~\label{sec:automorphism}
Automorphism is a process that permutes the coefficients of a polynomial by using Galois group. \Design supports automorphism based on the observation from BTS~\cite{bts}: the interleaved coefficients (Section~\ref{sec:bank_layout}) in the same tile (mat in \Design) will be mapped to a single tile after automorphism. \Design further extends this idea to interleave coefficients in one more dimension, memory row, where the column $c$ of row $z$ of a mat $(x,y)$ stores coefficients with the indices $cN_xN_yN_z+zN_xN_y+yN_x+x$. With such coefficient mapping, the automorphism only requires three steps: permutations in each row, vertical inter-mat permutation, and horizontal inter-mat permutation. \Design can handle the first step in NMU and the last two steps using MDLs and HDLs respectively.
\subsection{Application Mapping Framework for \Design}~\label{sec:mapping}
\Design can provide large throughput when fully utilizing memory. However, it is not trivial to map a full FHE program to \Design hardware with high utilization. 
Thus, we propose a mapping framework that generates data layout and scheduling in a pipeline manner that can fully utilize the memory to process multiple input data in parallel. 

\begin{figure}[t!]
    \centering
    \epsfig{file=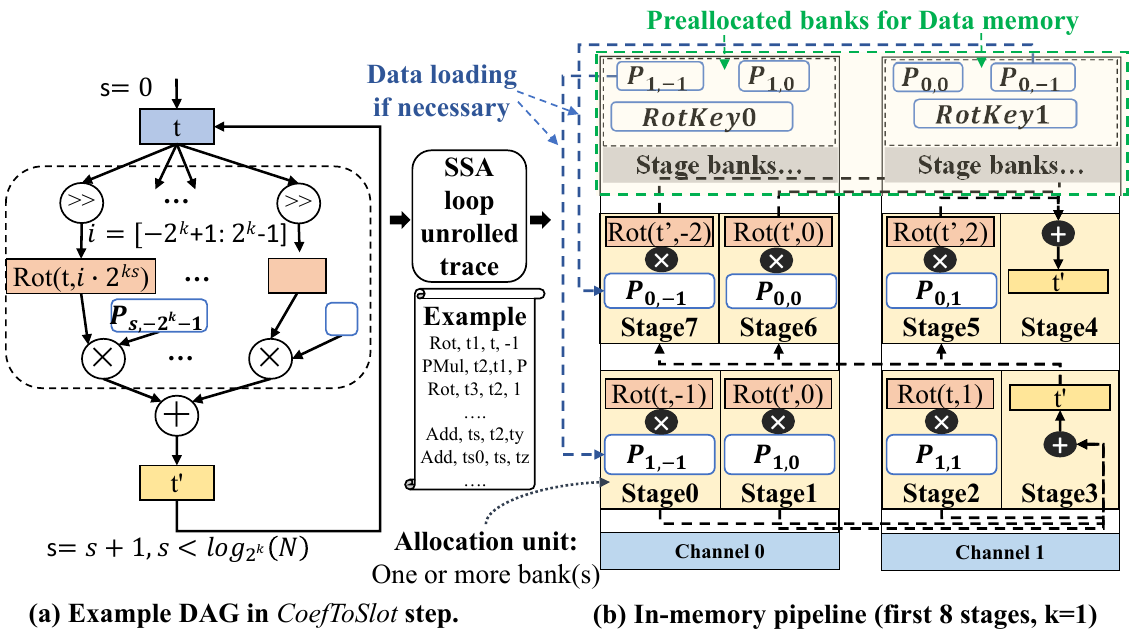, width=\columnwidth}
    \squeezeup
    \squeezeup
    \caption{In-memory pipeline generation.} 
    \squeezeup
    \label{fig:pipeline_mapping}
\end{figure}

\subsubsection{Framework Overview}
Figure~\ref{fig:pipeline_mapping} shows an example pipeline for the CoefToSlot step in CKKS bootstrapping. The input of our framework is an intermediate representation extracted from the real FHE program. Our framework generates a trace of FHE operations (e.g., HMul, HAdd, and HRot) in the static single-assignment (SSA) form while unrolling all loops. Our framework then divides the operation trace into multiple pipeline stages. The example shows the first 8 pipeline stages for CoefToSlot in a simplified HBM model with 2 channels. After computation on a stage, the allocated memory needs to transfer data to the memory that processes the pipeline steps with a data dependency. Therefore, the latency of each pipeline stage includes loading time, computation time, and transfer time. Our framework aims to minimize the latency of the bottleneck stage in the pipeline.

\subsubsection{Memory Allocation for Pipeline Stages}
Our framework allocates each stage to a basic allocation memory unit, whose size is determined by the FHE parameter setting including the polynomial degree, ciphertext scaling factor, and ciphertext moduli. In Figure~\ref{fig:pipeline_mapping}, the basic memory unit is one bank. To process a stage, we need sufficient memory to support the data layout shown in Section~\ref{sec:bank_layout} for the input and output data. Extra memory is needed for constant data (e.g., evk and plaintexts). The ideal case is each allocation unit can hold all data for the stage. When a memory allocation unit cannot hold all data, we store constant data in a reserved memory location called data memory. When data is stored in the data memory, all operations (across all stages) requiring the data need to dynamically load it. We reduce the memory footprint by storing data in one place, instead of duplicating them in different memory locations.

\begin{figure}[t!]
    \centering
    \epsfig{file=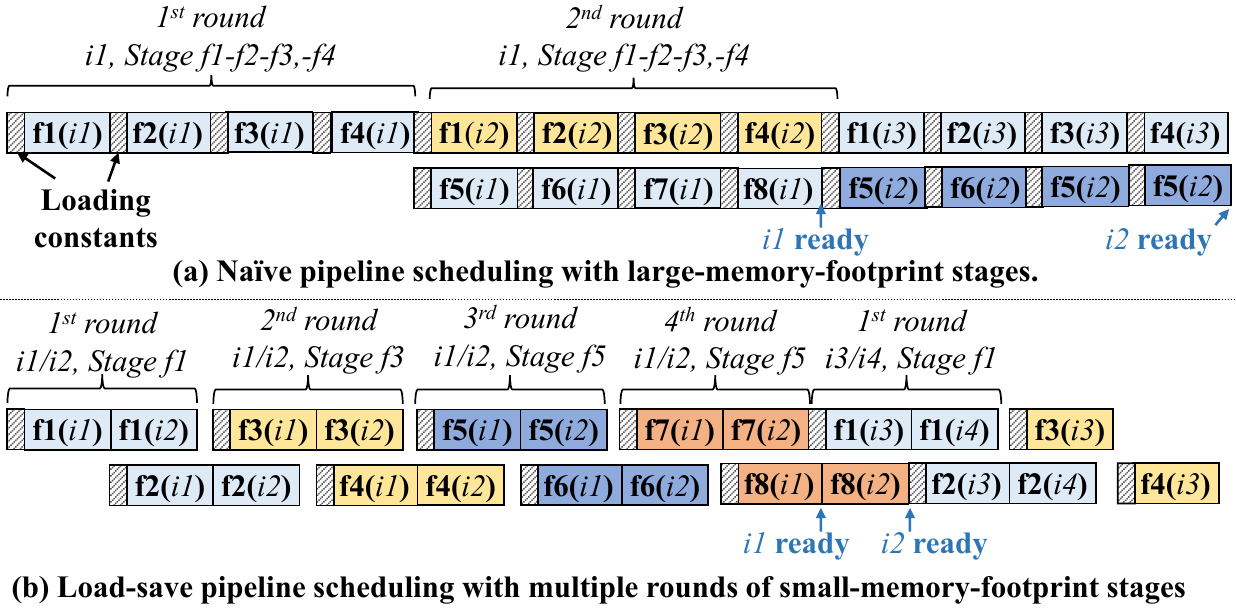, width=\columnwidth}
    \squeezeup
    \squeezeup
    \caption{Load-save pipeline optimization in two memory partitions.} 
    \squeezeup
    \label{fig:pipeline_optimization}
\end{figure}
\subsubsection{Load-save Pipeline}
A naive way of pipeline generation is to divide the FHE program into $n$ stages, where $n$ is the number of available allocation units in memory. However, for large applications, each stage may require a large memory footprint for operations, leading to frequent constant loading. As shown in Figure~\ref{fig:pipeline_optimization}(a), a stage has 4 operations, and every operation needs to load constants because memory is occupied by the constants of previous operations. In \Design, we propose a load-save pipeline by dividing operations into fine-grained stages with a small enough footprint. The fine-grained stages are allocated to different memory in a round-robin manner, requiring multiple rounds to process all stages for an input. In each round, the memory only creates a pipeline with part of the program (f1 and f2 in the first round in Figure~\ref{fig:pipeline_optimization}(b)). It runs through a batch of input with only 1 data loading at the beginning of each round. Each memory loads the next round of stages when the current input batch is completed. The load-save pipeline minimizes the data loading while still fully utilizing the memory for computation. 
\section{Experimental Setup}\label{sec:exp}

\begin{table}[t!]
  \centering
  \caption{Architectural parameters.}
  \squeezeupless
  \label{tab:sys_config}
\resizebox{\columnwidth}{!}{
\begin{tabular}{|l|l|}
\hline
\textbf{HBM configuration}  &  8-high HBM2E (16GB/stack)@10nm       \\ \hline
\textbf{Memory organization} & \#banks/pseudo-channel=8, \#pseudo-channels/stack=32 \\  \hline
\textbf{Bank specification}          & 64MB, row\_size=1kB, 512*512 mats\\ \hline
\textbf{Data transfer} & inter-bank NoC = 256-bit    \\  \hline
\textbf{Timing (ARx1)} & $tRRD:2ns$, $tRAS:29ns$, $tRP:16ns$, $tFAW:12ns$  \\ \hline
\textbf{Energy @10nm (ARx1)} & $row\_act:413pJ$, $pre\_gsa:0.69pJ/b$\\
& $post\_gsa:0.53pJ/b$, $IO:0.77pJ/b$\\ \hline
\end{tabular}
}
\squeezeup
\end{table}

\squeezeupless

\subsection{Hardware Evaluation}
\textbf{Memory Technology:}\label{sec:mem_valid}
The basic architecture of \Design is similar to HBM2E~\cite{8gbhbm,jedechbm}.
The system configuration is shown in Table~\ref{tab:sys_config}. Each HBM2E stack has 16GB with 16 physical channels. We scale energy and power values from 22nm used in previous work~\cite{o2017fine} to 10nm (shown in Table~\ref{tab:sys_config}), based on the recent HBM2E technology~\cite{8gbhbm}. We follow the method of Vogelsang~\cite{mem_scaling} to calculate the scaling factors for energy. 
We assume 16 physical channels on a stack are connected by a crossbar on PHY where each bidirectional link is 64-bit wide (bisection bandwidth=64GB/s per stack). We also add stack-stack links for scaled-up systems, commonly used in memory-centric architecture~\cite{giannoula2021syncron}. \Design has two HBM2E stacks to support 32GB memory. We exploit the remaining signaling links on HBM2E for stack-stack connection so the inter-stack bandwidth is also 256GB/s.

\textbf{Hardware Modeling:}\label{sec:hw_valid}
To evaluate the area and power of customized components, we synthesize our design in 45nm technology using Nangate Open Cell Library. We model all other CMOS components (including buffers and interconnects) in Cacti~\cite{cacti} at 32nm technology.
We scale all values to the 10nm technology with the scaling factors calculated from previous work~\cite{scaling}. We estimate the delay, power, and area overhead of integrating CMOS-ASIC and DRAM technologies based on the difference in number of metal layers and complexity of the customized logic~\cite{drisa}. 

\textbf{Simulation:}
We generate FHE operation traces from software implementations of different workloads, and our mapping framework optimizes the trace and generates PIM instructions for simulation. Our in-house simulator adopts a cycle-accurate trace simulation based on the standardized DRAM latency constraints, similar to Ramulator~\cite{ramulator}. We model control logic at different levels in the DRAM hierarchy.

\begin{figure*}[t!]
    \centering
    \epsfig{file=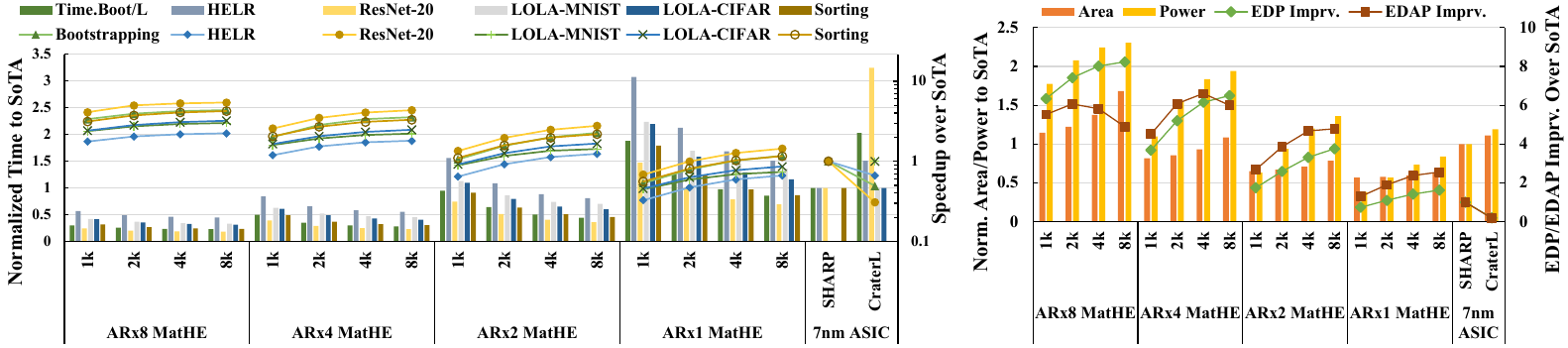, width=\textwidth}
    \squeezeup
    \squeezeup
    \caption{Efficiency of different \Design configurations. Deep/shallow workloads are normalized to SHARP~\cite{sharp}/CraterLake~\cite{craterlake}.}
    \squeezeup
    \label{fig:overall_comp}
\end{figure*}

\subsection{Workloads}
\textbf{Logistic Regression (HELR)~\cite{han2019logistic}:}
This workload has 30 iterations of homomorphic logistic regression where each iteration trains 1024 samples with 256 features as a batch. The multiplication depth is deep, requiring several bootstrappings.

\textbf{ResNet-20~\cite{resnet}:}
The ResNet-20 is a homomorphic neural network inference for one CIFAR-10 image classification. The network is deep with multi-channel convolutions, matrix multiplications, and approximated ReLU function.

\textbf{Sorting~\cite{sorting}:}
Sorting uses 2-way bitonic sorting on an array with 16,384 elements, the same as that used in SHARP~\cite{sharp}.

\textbf{Bootstrapping~\cite{ckks_boot}:}
We evaluate the bootstrapping algorithm using a similar framework as previous work~\cite{ckks_boot,ark}, which requires 15/30 levels of bootstrapping. We adopt the minimum-key method used in previous work~\cite{ark} to reduce the rotation keys used in bootstrapping.

\textbf{Shallow neural network inference (LOLA)~\cite{lola}:}
We also evaluate two shallow workloads without bootstrapping, including a network for MNIST (LOLA-MNIST) and a larger network for CIFAR-10 (LOLA-CIFAR), used in CraterLake~\cite{craterlake}.

\subsection{FHE Parameters and Evaluation}~\label{sec:fhe_params}

We evaluate the efficiency of \Design on a 128-bit security FHE parameter setting chosen from Lattigo~\cite{lattigo}. 
For workloads with bootstrapping, including HELR, ResNet-20, and sorting, we use $logN=16$, $L=23$, $dnum=4$, and $logPQ=1556$, similar to prior accelerators~\cite{bts,ark}. Each polynomial is decomposed into 40-61 bit RNS terms, where \Design allocates 64-bit for each coefficient. For shallow LOLA workloads, we use the similar parameter settings with CraterLake~\cite{craterlake}, where we assume $logN=14$, $L=4/6$, and $logq_i<=32$. \Design represents 32-bit coefficients in 64-bit words and packs 4 RNS $logN=14$ polynomials together in 16 subarrays.
For all workloads, we measure the maximum time across all pipeline stages, indicating the time we can finish an input when the pipeline is full.
In addition, we consider the number of pipelines that can be processed simultaneously in the system when the program cannot fully utilize the memory capacity (e.g., 32GB).

\section{Evaluation}

\subsection{Comparison to Previous FHE Accelerators} \label{sec:soa_comp}
Figire~\ref{fig:overall_comp} compares \Design with two state-of-the-art ASIC FHE accelerators (CraterLake~\cite{craterlake}, and SHARP~\cite{sharp}). We explore two design choices of memory organization that play important roles in the performance, power, and area of \Design: the aspect ratio of DRAM mat (AR), and the width of adders in a subarray (if each NMU has 2 64-bit adders, a subarray with 16 mats has 2k-width adders). 
As discussed in Section~\ref{sec:pim_motiv}, high-AR architecture has better performance and energy efficiency than low-AR architecture, but incurs significant area overhead. The width of adders determines the performance of arithmetic computing - wide adder designs support faster computing while requiring a larger area than narrow adders. 
For the chip area of prior ASIC accelerators, we add the area of 32GB HBM2E (2$\times$110$mm^2$) for a fair comparison. 

\subsubsection{Performance}
\Design shows superior performance over prior FHE accelerators. Specifically, ARx8-8k (lowest EDP) is 4.4$\times$ (8.8$\times$), 2.2$\times$ (3.4$\times$), and 5.4$\times$ (17.5$\times$) faster than SHARP~\cite{sharp} (CraterLake~\cite{craterlake}) on bootstrapping, HELR, and ResNet-20 respectively. ARx4-4k (lowest EDAP) is 3.4$\times$ (6.8$\times$), 1.7$\times$ (2.6$\times$), and 4.1$\times$ (13.2$\times$) faster than SHARP~\cite{sharp} (CraterLake~\cite{craterlake}) on bootstrapping, HELR, and ResNet-20 respectively. For sorting, ARx8-8k (ARx4-4k) is 4.2$\times$ (3.1$\times$) faster than SHARP~\cite{sharp}. For LOLA-MNIST and LOLA-CIFAR, ARx8-8k (ARx4-4k) are 3.0$\times$ (2.1$\times$) and 3.2$\times$ (2.3$\times$) faster than CraterLake~\cite{craterlake}.
The less significant performance improvement in HELR results from the low portion of bootstrapping which is significantly optimized by adopting the minimum-key optimization~\cite{ark,sharp}.

\subsubsection{Efficiency}
We then compare the power and area efficiency of \Design and ASIC accelerators. 
Compared to SHARP, ARx8-8k improves EDP and EDAP by 8.2$\times$ and 5.1$\times$. However, ARx8-8k requires 1.6$\times$ larger area and 2.3$\times$ higher power than SHARP. ARx4-4k improves EDP and EDAP of SHARP~\cite{sharp} by 6.2$\times$ and 6.9$\times$, with 0.9$\times$ area and 1.8$\times$ power consumption. ARx2-2k is a configuration that provides the best performance using less area (0.65$\times$) and power (0.94$\times$) than SHARP. For performance, ARx2-2k is 1.56$\times$, 0.92$\times$, and 1.96$\times$ faster than SHARP, leading to 2.59$\times$ and 3.96$\times$ EDP and EDAP improvement.

\subsubsection{Analysis of \Design Benefits}
Compared to ASIC accelerators, \Design provides a higher throughput due to the efficient in-memory computation and large intra-memory bandwidth. For instance, ARx4-4k \Design has 16 million 64-bit adders. Considering the cost of DRAM row activations, data transfers, subarray-level parallelism, and 500MHz frequency of additions, the effective throughput of ARx4-4k for 64-bit multiplication is around 637.61 TB/s. During multiplication, the adders consume most of the energy because row activation energy is amortized for the entire row, and data transfer is energy-efficient due to the short wire length. Therefore, the energy efficiency of \Design computation is similar to the modular multiplier used by FHE accelerators (slightly higher due to the DRAM-CMOS integration).
For the internal bandwidth of NTT, ARx4 \Design supports a 256-bit link (500MHz) for each of the 512 subarrays in a bank. Considering up to half of the subarrays can transfer data simultaneously during NTT, the peak internal bandwidth for NTT is 2048 TB/s in 32GB ARx4 \Design. For the slowest NTT step, the internal bandwidth drops by 16$\times$ (128 TB/s). As a comparison, SHARP~\cite{sharp} has around 24K 36-bit multipliers running at 1GHz, leading to 221.18 TB/s throughput. Furthermore, the on-chip memory resources in SHARP support 72TB/s bandwidth.

\subsection{Comparison across different \Design Configurations}
As shown in Figure~\ref{fig:overall_comp}, There is a significant difference in power and area between different \Design configurations. For example, the most oversized \Design (ARx8-8k) requires 642.32$mm^2$ chip area and 218W power, while the smallest \Design (ARx1-1k) only requires 223.81$mm^2$ chip area and 36.24W power. As a reference, the commercial 2-stack HBM2E has a chip area of 220$mm^2$~\cite{8gbhbm} and the power budget of a conventional HBM system is 60W~\cite{o2017fine}. We note that the power budget of conventional HBM is different from the accelerator design targeted in this work, where high power consumption is reasonable if it meets the thermal requirement (e.g., 10W/$cm^2$/layer~\cite{mem_temperature}).

Based on the results, high-AR \Design provides higher performance than low-AR designs because increasing AR can increase both compute and data movement throughput inside a bank. For ARx1 and ARx2 \Design configurations, doubling AR can provide 1.57$\times$ to 1.98$\times$ speedup because the execution is compute-bound. For ARx4, doubling AR only provides 1.23$\times$ to 1.67$\times$ speedup. Increasing adder-width exhibits a similar trend where the effect of increasing compute resources diminishes for high-throughput architectures.

To find the most cost-efficient \Design design, we evaluate energy-delay-product (EDP) and energy-delay-area-product (EDAP) for different \Design configurations. For EDP, the trend follows the performance, where the largest \Design (ARx8-8k) gives the lowest EDP. When considering the area cost, different ARs favor different adder widths. Specifically, ARx8 and ARx4 exhibit the lowest EDAP at 2k and 4k adder-width respectively. The configuration with the lowest EDAP (ARx4-4k) is 1.34$\times$ more efficient than ARx8-8k.

\subsection{\Design Latency and Energy Analysis}\label{sec:breakdown}
\begin{figure}[t!]
    \centering
    \epsfig{file=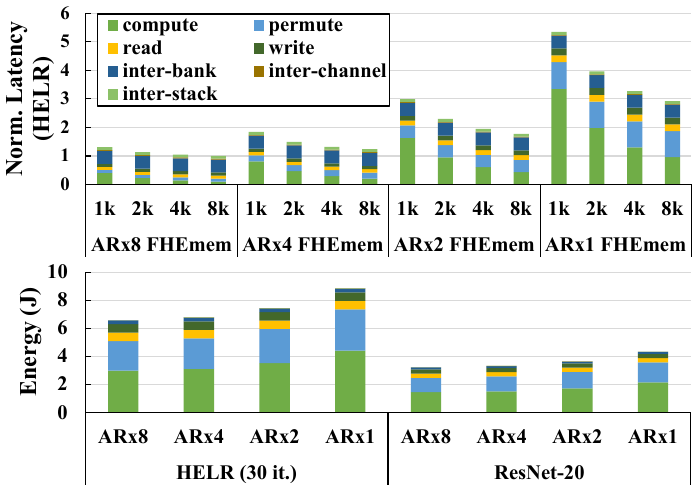, width=0.85\columnwidth}
        \squeezeup
    \caption{The latency and energy breakdown.}

    \squeezeup
    \label{fig:breakdown}
\end{figure}
Figure~\ref{fig:breakdown} shows the latency and energy breakdown of different \Design configurations. Considering the parallel processing, we accumulate all latency values across all memory banks for the latency breakdown. We divide all operations into 7 categories, including computation (subarray activation/precharge, operand transfer, and addition), permutation (inter-mat transfer), read/write (activation and precharge for data transfers), and inter-bank/channel/stack IO traffics. 

The breakdown results provide several key insights into \Design. First, in low-AR \Design, the latency is dominated by computation and permutation because of the limited throughput of computation and intra-bank data movements. Increasing AR can effectively reduce the latency for both computation and permutation latency. Furthermore, increasing the adder width can effectively reduce the computation latency. However, in high-AR \Design, the data movement becomes the performance-dominant operation, especially inter-bank data movements (mainly caused by BConv). This proves the necessity of customized inter-bank links. We analyze the detailed effect of optimizations in Section~\ref{sec:benefits}. For energy consumption, \Design consumes most of the energy on computation and permutation, which incurs intensive row-activation and intra-bank data movements.

\subsection{PIM Technologies}
\begin{figure}[t!]
    \centering
    \epsfig{file=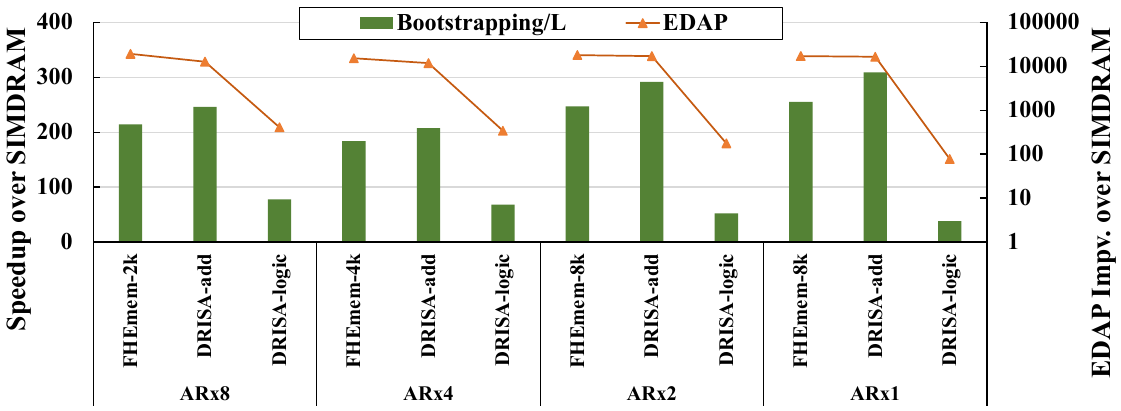, width=\columnwidth}
    \squeezeup
    \squeezeup
    \caption{Comparison between PIM technologies.}
    \squeezeup
    \label{fig:pim_tech}
\end{figure}
Figure~\ref{fig:pim_tech} compares the efficiency of \Design to other PIM technologies, including SIMDRAM~\cite{simdram} and DRISA~\cite{drisa}. For a fair comparison, we use the proposed application mapping and customized data links in baseline PIM architectures, while evaluating the difference in processing. We implement an adder-less NMU for permutations in baselines. We select the most efficient (lowest EDPA) \Design for each AR. The results show \Design is 183.7$\times$ to 255.4$\times$ faster than SIMDRAM~\cite{simdram}, and 2.76$\times$ to 6.75$\times$ faster than DRISA-logic~\cite{drisa}. Furthermore, \Design is at least 19,300$\times$ and 47$\times$ more efficient than SIMDRAM and DRISA-logic using EDAP. Compared to DRISA-add~\cite{drisa}, \Design is 1.14$\times$ to 1.21$\times$ slower from ARx8 to ARx1 because DRISA's adders can directly access the sense amplifiers. However, \Design is 1.04$\times$ (ARx1) to 1.51$\times$ (ARx8) more efficient in EDAP because \Design does not introduce area overhead in a mat. Furthermore, DRISA is more challenging to manufacture because the large adder area will affect the alignment of cost-optimized bitlines. As a comparison, \Design puts all customized logic outside the mat structure.

\subsection{Overhead Analysis}
\begin{table}[t!]
  \centering
  \caption{Area and power of \Design (16GB HBM2E).}
  \label{tab:area_power}
\resizebox{\columnwidth}{!}{
\begin{tabular}{l|llll}
\hline
\textbf{ARx4 HBM }      & \textbf{DRAM Cell}  & \textbf{Local WL Driver} & \textbf{Sense Amp}    & \textbf{Row/Col Dec.s} \\ \hline
\textbf{Area ($mm^2$)}     & 56.54      & 26.15           & 45.63        & 0.39     \\ \cline{2-5} 
               & \textbf{Center Bus} & \textbf{Data Bus}        & \textbf{TSV}          & \textbf{Total}    \\ \cline{2-5} 
               & 1.56       & 4.81            & 13.25        & 148.33      \\ \hline
\textbf{4K adder}   & \textbf{Horizontal DL}     & \textbf{Adder\&Latches}           &\textbf{Bank Chain \& Buf.}   &   \textbf{Control logic} \\ \hline
\textbf{Area ($mm^2$)}     & 14.13      & 30.43            &   0.065    &  0.56$mm^2$   \\
\textbf{Power/Energy} & 5.3fJ/b (avg.)      & 15.86W         &   0.53pJ/b     &   0.12W   \\ \hline
\end{tabular}
}
\squeezeup
\end{table}
Table~\ref{tab:area_power} shows the area and power breakdown for our customized hardware components of \Design based on 1 HBM2E stack (16GB). We exploit the Cacti-7~\cite{cacti} to generate the area breakdown of HBM and rescale the values to the published work~\cite{8gbhbm}. The table shows a structure of ARx4 HBM and each subarray contains 4k-wide adders. All area values are for a single layer. The large area mainly comes from the near-mat adders. The horizontal data links use the same technology as the global data lines with the consideration of energy efficiency ($4\times$ larger than local bit-lines). We extract the capacitance of material and scaling method from previous work~\cite{mem_scaling} to calculate the energy consumed by data transfer. We note that the target of \Design is not a cost-optimized memory product, but a specialized accelerator for emerging applications that require high-throughput computation and/or efficient intra-memory data movements. However, the proposed design is still practical regarding area and power overhead. First, unlike previous near-subarray PIM~\cite{drisa}, \Design put the customized logic outside DRAM mat, avoiding the issue of aligning cost-efficient local bitlines. Second, \Design can maintain an average DRAM working temperature (under 85$^{\circ}C$). Previous work~\cite{mem_temperature} shows a 16-high compute-centric 3D memory can tolerate 10W/$cm^2$ per memory layer to keep DRAM temperature under the 85$^{\circ}C$ limit with a commodity-server active heat sink. For example, the power consumption and area of 8-high ARx4 4k \Design are 173.9W and 367$mm^2$, resulting in a power density of 5.92W/$cm^2$ (the highest power density in our exploration).

\subsection{Evaluation of \Design Optimizations} \label{sec:benefits}
\begin{figure}[t!]
    \centering
    \epsfig{file=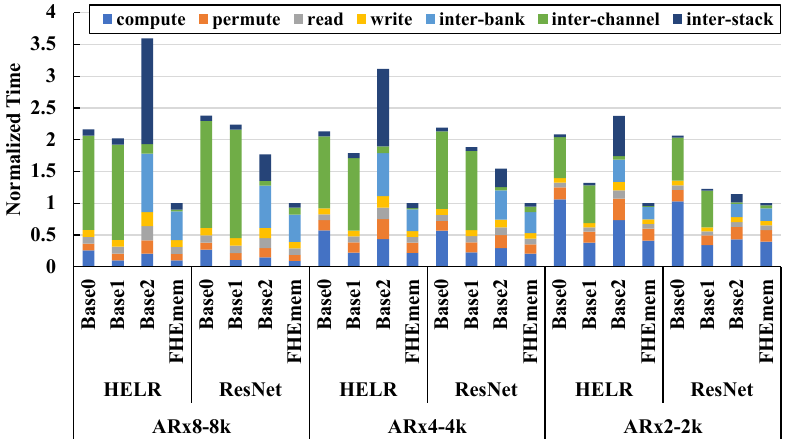, width=\columnwidth}
    \caption{Effect of different optimizations including (1) Montgomery-friendly moduli, (2) inter-bank connection network, and (3) load-save pipeline mapping. Base0 uses (3), Base1 uses (1)+(3), and Base2 uses (1)+(2).}
    \squeezeup
    \label{fig:sensitive}
\end{figure}

We compare \Design with baseline systems enabling a subset of \Design optimizations, as shown in Figure~\ref{fig:sensitive}.

\subsubsection{Montgomery-friendly Moduli}
The Montgomery-friendly moduli can reduce the number of addition steps, improving the computation performance. As shown in Figure~\ref{fig:sensitive}, Base1 is 1.68$\times$ and 1.58$\times$ faster than Base0 on HELR and ResNet with ARx2-2k architecture. On Arx4 and Arx8 architectures, this optimization only improves the performance by 1.17$\times$ and 1.06$\times$. However, Montgomery-friendly moduli can reduce energy consumption by 1.75$\times$ because computation is energy-dominant across all \Design architectures (Figure~\ref{fig:breakdown}).

\subsubsection{Interconnect Network}
The comparison between \Design and Base1 shows the efficiency of the proposed inter-bank network, where Base1 uses the existing HBM channel IO for all inter-bank data movements. Based on our results, the proposed inter-bank network can improve the performance by 1.31$\times$, 1.86$\times$, and 2.12$\times$ on ARx2, ARx4, and ARx8 respectively. The inter-bank network can reduce the latency of related data movement by 3.2$\times$ on average.

\subsubsection{Load-save Pipeline Mapping}
The comparison between \Design and Base2 shows the efficiency of the load-save pipeline mapping. For HELR, load-save pipeline mapping improves the performance by 3.59$\times$ (1.77$\times$), 3.12$\times$ (1.54$\times$), and 2.38$\times$ (1.15$\times$) on ARx8, ARx4, and ARx2, respectively, for HELR (ResNet). The significant performance improvement results from reducing frequent data loading, especially data from a remote stack.
\section{Related Work}

Several FHE-specific accelerators have been proposed recently in the architecture community~\cite{cryptopim,craterlake,bts,ark,sharp}. 
HEAX~\cite{heax} is an FPGA-based accelerator targeting the CKKS FHE scheme, and accelerates FHE primitives (e.g., multiplication and key-switching) by up to 200$\times$; HEAWS~\cite{heaws} is also an FPGA implementation on Amazon AWS FGPAs, which accelerates B/FV FHE scheme (5$\times$ on a microbench). 
CraterLake~\cite{craterlake} adopts wide vector processor with specialized high-throughput units for \texttt{BConv} and on-chip key generation. BTS~\cite{bts} exploits relatively low throughput function units with large inter-PE crossbar network. ARK~\cite{ark} uses an algorithm-hardware co-design to significantly reduce the off-chip bandwidth for bootstrapping $evk$ and plaintext polynomials. SHARP~\cite{sharp} further improves the performance ARK by using low bit-precision data path (36-bit vs. 64-bit in ARK). However, their technique does not apply to general $evk$ and ciphertexts, leading to the same memory issues, as analyzed in Section~\ref{sec:motivation}.
\Design exploits the large internal memory bandwidth of memory-based acceleration to unleash the processing throughput in a more area and power-efficient way.

CryptoPIM~\cite{cryptopim} is a ReRAM-based PIM accelerator with customized interconnect for NTT operations, lacking the support for more general FHE operations. In-storage processing is another promising technology to accelerate big-data applications~\cite{inspire}. INSPIRE is an in-storage processing system for private database queries based on FHE by integrating FHE logic (e.g., NTT and permutation) in the SSD channels. INSPIRE only supports small FHE parameters (i.e., N=4096) and its throughput is limited by the number of SSD channels. MemFHE~\cite{memfhe} is a ReRAM-based PIM accelerator with the customized data path for bit-level TFHE scheme, not applicable to CKKS and other packed FHE schemes.
\section{Conclusion}
This work proposes \Design, an FHE accelerator based on a novel near-mat processing architecture with relatively lightweight hardware modifications in conventional HBM. We also propose the end-to-end processing flow with a mapping framework for PIM-based FHE. Our evaluation shows \Design is 6.9$\times$ more efficient than prior-art ASIC accelerator.

\section*{Acknowledgement}
This work is supported by SRC HWS, SRC PRISM, and NSF fundings \#2003279, \#1911095, \#1826967, \#2100237, \#2112167, \#2052809, \#2112665. 

\bibliographystyle{IEEEtran}
\bibliography{refs}

\vfill

\end{document}